\documentstyle[12pt]{article}
\date{}
\begin{document}
\centerline{\bf EF\/FECTIVE FERMION MODELS IN SYMMETRY-BREAKING}
\centerline{\bf PHASE AND QUANTUM CHROMODYNAMICS
\footnote{The talk given at the International School-Seminar '93 -
Hadrons and Nuclei from QCD (21 Aug.-3 Sept. 1993, Vladivostok-Sapporo}}
\hfill\\
\hfill\\
\centerline{A. A. ANDRIANOV and V. A. ANDRIANOV}
\centerline{\footnotesize\it Department of Theoretical Physics,}
\centerline{\footnotesize\it University of Sankt-Petersburg,}
\centerline{\footnotesize\it 198904 Sankt-Petersburg, Russia}
\hfill\\
\begin{abstract}
In our lecture we discuss
the fermion models with quasilocal interaction implemented by derivatives
and a momentum cutof\/f as substitutes of QCD at low energies.
They are investigated in the strong coupling regime
when several coupling constants
are matched to their critical values. It is found that around polycritical
points there appear a number of resonances with the same quantum numbers.
Respectively the particular change of enviroment caused by gluon condensate
results in the mass
splitting independent of the cutof\/f. Such models are supposed to be
an essential ingredient in the description of quark matter at high baryon
densities.
\end{abstract}
\section*{\normalsize\bf 1. Introduction}
\hspace*{3ex} The ef\/fective quark models with four-fermion interaction are
widely used as substitutes of the low-energy QCD in the hadronization
regime$^{1,2,3}$. They are applied also to describe the minimal extensions
of the Standard Model (Top-mode SM$^{4,5,6}$). At strong coupling they reveal
the dynamic breaking of chiral symmetry (DCSB) which is essential
in description of light hadrons. As well as in QCD in the quark models
DCSB arises due to fermion condensation $\langle \bar\psi\psi\rangle
\not= 0$ that corresponds to the creation of dynamic quark mass $m_{dyn}
\not= 0$

In the common approach the local
four-fermion interaction only is employed$^{7,9}$. Meanwhile
in these models very complicated gluon forces between quarks are
replaced by a color-singlet self-interaction of quarks. Such an ef\/fective
action should generally
contain  quasilocal vertices with derivatives$^{6,10}$. Moreover as we
will see the simplest quark model does not take the main properties of QCD
which are important to reproduce the variety of hadron states.
Namely an ef\/fective quark model that reflects reasonably QCD-born mass
spectrum at low and intermediate energies should obey the following
requirements:
\begin{itemize}
\item[(i)] at strong couplings the DCSB should arise and the expected set of
hadron states should be generated by the universal DCSB
(including radial excitations with the same quantum numbers,
$\pi,\,\pi',\,\pi'';\quad \sigma,\,\sigma',\,\sigma''$ etc., see Review of
Particle Properties);
\item[(ii)] both DCSB and hadron state formation should be derived in the
large-$N_c$
approach, i.e. in the mean-field approximation for fermion fields;
\item[(iii)] the mass splitting of hadron states should be induced by a single
force in order to provide the universal mass scale
$m_{hadron} \sim \Lambda_{QCD} \sim m_{dyn}$.
\end{itemize}

We are going to show that above requirements can be realized in fermion
models with quasilocal vertices with derivatives. In order to generate
"radial excitations" one has to admit strong (critical)
coupling constants for them. Thereby we come to the generalized
Nambu-Jona-Lasinio (NJL) model in the vicinity of polycritical point for
several coupling constants.

There are two  important reasons to make this generalization:
\begin{enumerate}
\item[(i)] it extends the energy range of applicability of such a quark
model and improves its accuracy at low energies;
\item[(ii)] such models are thought of as more realistic for investigations
of hadron matter at high temperatures and nuclear densities near the
restoration of CSB and deconfinement.
\end{enumerate}

\section*{\normalsize\bf 2. Dynamic Symmetry Breaking in Models with
Four-fermion Interaction \hspace*{3ex}and the Critical Point (Fine-tuning)}
\hspace*{3ex}Let us remind how the DCSB arises in a model
with local 4-fermion interaction due to strong attraction in the scalar
channel. The simplest, Gross-Neveu model retaining the scalar channel only can
be presented by the lagrangian density,
in two forms (the euclidean-space formulation is taken here),
\begin{equation}
{\cal L}\,= \,\bar\psi\not\!\! D \psi\,+\,\frac{g^{2}}{4N_{c}}\,
(\bar\psi \psi)^{2}
\,=\,\bar\psi\bigl(\not\!\! D\,+\,i m (x)\bigr)\psi\,
+\frac{N_{c}}{g^2}\,m^{2}(x),
\end{equation}
where $ \not\!\! D \,=\,i\gamma_{\mu}\partial_{\mu}$ and $\psi \equiv \psi_{i}$
stands for color fermion f\/ields with $N_{c}$ components.
We simplify our analysis and set up the
number of f\/lavours
$N_{F}=1$ and the current quark mass $m_{q}=0$. In Eq.1
the scalar auxiliary f\/ield $m (x)$ is introduced in order to describe
the CSB phenomenon in the large-$N_{c}$ limit.

This model is implemented by a cutof\/f $\Lambda$ for fermion energy spectrum.
Namely we define low-energy fermion fields by means of the projection
$\,\psi_{l}=\Theta(\Lambda^2 -
\not\!\! D^2)\psi$ where $\Theta(x)$ is a step-function$^{12}$ (or any other
regulator). For a quark model the cutof\/f $\Lambda$ can be thought of as
 a separation scale which appears when constructing the QCD low-energy
ef\/fective  action. From this viewpoint the observables should not depend
on the latter one. The scale invariance is achieved by appropriate
prescription of cutof\/f dependence for ef\/fective coupling constants $g$.

The regularized ef\/fective action $S_{eff}$ for auxiliary field,
\begin{equation}
Z_{F}^{\Lambda}(m)\,=\, \exp(- S_{eff})\,=\,\biggl\langle \exp\biggl(
-\int\! d^{4}x\,{\cal L}_{F}(m (x))\biggr)\biggr\rangle_
{\bar\psi\psi}
\end{equation}
possesses the mean-field extremum on constant configurations $m = const$.

The relevant ef\/fective
potential $V_{eff}$ can be obtained by intergration over fermions,
\begin{equation}
V_{ef\/f}\,=\, \frac{S_{eff}}{(vol.)} \,=
\,\frac{N_{c}}{8\pi^2}\biggl\lbrace\frac{\Lambda^4}{2}\bigl({1\over 2} -
\ln\frac{\Lambda^2 + m^2}{\mu^2}\bigr)
 - \frac{m^2\Lambda^2}{2}
+ \frac{m^4}{2}\ln\frac{\Lambda^2 + m^2}{m^2} +
\frac{8\pi^{2} m^2}{g^2}
\biggr\rbrace \, .
\end{equation}
Its extrema are arrived from the
mass-gap equation,
\begin{equation}
R(m)\, \equiv\,\frac{4\pi^2}{N_{c}}\cdot
\frac{\partial V_{ef\/f}}{\partial m}
= m \,\biggl(\frac{8\pi^2}{g^2}\,+\,m^2
\ln\frac{\Lambda^2 + m^2}{m^2}
\,- \,\Lambda^2\biggr)\,=\,0 .
\end{equation}
The main contribution into Eq.4 is given by a tadpole term in the
one-fermion loop which is related to v.e.v of scalar fermion operator,
\begin{equation}
R( m ) = \,m \frac{8\pi^2}{g^2}\, +\, i\frac{4\pi^2}{N_{c}}
\langle \bar\psi \psi \rangle \, .
\end{equation}
The cutof\/f independence is realized with aid of fine-tuning$^8$,
$\frac{8\pi^2}{g^2} = \Lambda^{2}$. In the language of the theory of
critical phenomena it is equivalent to developing of our model around
critical or scaling point. By definition the critical coupling constant
is $g_{crit}^2 = 8\pi^2/\Lambda^2$. When  $g^2 < g_{crit}^2$ the only
solution of mass-gap Eq.4 is $m = 0$, while for  $g^2 > g_{crit}^2$ there
exists a nontrivial dynamic mass solution $m_{dyn} \not= 0$ which brings
the true minimum for $V_{eff}$.

The f\/ine-tuning states that the strong
 $\Lambda^2$-dependence should be compensated by the corresponding term
in the coupling constant,
\begin{equation}
\frac{8\pi^2}{g^2} = \Lambda^{2} - m_0^2
\end{equation}
The deviation scale $m_0^2$ determines the physical mass of scalar meson.
Namely its kinetic term  can be obtained from the second
variation of $S_{eff}$,
\begin{eqnarray}
S_{eff} &\,\simeq\,& S_{eff}(m=m_{dyn})\,+\,\frac{1}{2}\int
\frac{d^4 p}{(2\pi)^4}\,
\sigma(-p) \,f^2_{\sigma}\, [p^2 + m^2_{\sigma}]\, \sigma (p);\nonumber\\
m &\,=\,& m_{dyn}\,+\,\sigma .
\end{eqnarray}
The scalar meson mass is given by the remarkable
Nambu relation $m_{\sigma} \simeq 2 m_{dyn}$
Respectively the scale $m_0^2$ should be weakly dependent
$m_0^2\,\sim\, m_{dyn}^{2}\,\ln({\Lambda^2}/ m_{dyn}^2)$
on the cutof\/f $\Lambda$ in order that
the physical mass parameters were decoupled on $\Lambda$,
$\partial_{\Lambda} m_{dyn} \,=\,0$.

What have we learned from this model?\\
\begin{enumerate}
\item[(i)] As a result of DSB in this model only one scalar meson is created
in the large-$N_c$ approach.
\item[(ii)] In such a simple model the radial excitations are not present in
the
large-$N_c$ approach.
\item[(iii)] The mass scale of scalar state is assumed to be dif\/ferent from
$\Lambda$ and related to the basic QCD scale $\Lambda_{QCD}$.
\item[(iv)] Still the hadron coupling constants, e. g. $f_{\pi},\,f_{\sigma}$,
remain weakly depending on $\Lambda$ in accordance to the $\log$-divergence
of self-energy diagram,$f^2_{\pi,\sigma} \sim \ln \Lambda^2/m^2_{dyn}$.
\item[(v)] Due to the fine-tuning the fermion condensate has a strong $\Lambda$
dependence,
\begin{equation}
\langle \bar\psi \psi\rangle\, \sim\, N_{c} m \Lambda^{2} \, .
\end{equation}
\end{enumerate}
Hence the conventional NJL (or GN)quark models do not contain a consistent part
of QCD ef\/fective action and at best can be used as a truncation$^3$
of the latter one at low energies or should be extended with inclusion of
higher dimensional vertices which are needed to cure enumerated defects$^{10}$.

In order to simulate the mass-splitting by QCD forces one can start
from a quark model pecisely at the critical point $g^2 = g^2_{crit},\,
m_{dyn} = 0,\, m_{hadron} = 0$ and drive the mass spectrum by coupling to
gluon medium (gluon condensates)$^3$. This scenario can be realized in a
Gauged NJL (or GN) model around critical point.

Let us consider its particular GN version in the gluon condensate expansion.
We replace,
\begin{equation}
\not\!\partial \Rightarrow \not\!{\cal D} = \not\!\partial + \not\! G,\quad
G_{\mu} = ig\,G^a_{\mu}\,t^a,
\end{equation}
with low-energy gluon fields saturating condensates.

In the large mass limit $V_{eff}$ is supplemented with
\begin{equation}
\Delta\,V_{eff}(G)\,=\,\frac{1}{48\pi^2} \ln\frac{m^2}{\Lambda^2}
\langle tr(C^a_{\mu\nu})^2 \rangle_0 \,=\,- \frac{C_g}{24}
\ln\frac{m^2}{\Lambda^2},
\end{equation}
where $C_g = (\alpha_s /\pi) \langle tr(C^a_{\mu\nu})^2 \rangle_0
\simeq (350\div400 MeV)^4$ is a gluon condensate.

The modified mass-gap Eq. is,
\begin{equation}
\frac{8\pi^2}{g^2}
\,- \,\Lambda^2\,+\,m^2
\ln\frac{\Lambda^2 + m^2}{m^2}\,=\, \frac{C_g\pi^2}{3N_c m^2}\,\equiv\,
\frac{\gamma m^2}{2} .
\end{equation}
When one drives to the overcritical region the positive gluon condesate
is assumed to induce the mass  $m_{dyn} \simeq 300 MeV$ obeying the Eq.
$\ln (\Lambda^2/m^2) \simeq \gamma/2$. Then  one has $\gamma = 4 \div 6$
and consequently $\Lambda \simeq (0.8 \div 1.0) GeV$.

Thus numerical estimations seem to support the concept of GNJL model with
critical four-quark constant. However the calculation of scalar meson decay
constant $f_{\sigma}$ shows too strong sensitivity to values of gluon
condensate,
\begin{equation}
f^2_{\sigma}\,\simeq\,\frac{N_c m^2_d}{8\pi^2}\biggl(
- \,\frac{17}{12}\,+\,\frac{3\gamma}{20}\biggr)_{\ln\frac{\Lambda^2}{m^2_d}
 = \frac{\gamma}{2}},
\end{equation}
i. e. $f_{\sigma}$ is negative for $\gamma = 4\div 6$. It makes such a model
unrealistic.

We will see later on that the two-channel generalization of GN model fits
noticeably better the decay constants and is less sensitive to the values of
$\gamma$.
\section*{\normalsize\bf 3. Dominant
Ef\/fective Vertices in the DSB regime and Generalization of
\hspace*{3ex}Gross-Neveu Model}
\hspace*{3ex}We consider the DSB pattern in the mean-f\/ield approach
(large-$N_{c}$ limit)
and estimate the vertices with any number of fermion legs and derivatives.
The main rule to select out relevant vertices is derived from the
requirement of indif\/ference in choice of separation
scale $\Lambda$ following the conception of low-energy effective action.

We assume that:
\begin{itemize}
\item[(i)] $\Lambda^2$-order vertices are dominant in creating the DSB-critical
surface that is achieved when all contributions of $\Lambda^2$-order
are cancelled;
\item[(ii)] $\Lambda^0$-orders in vertices assemble in the mean-f\/ield action
to supply fermions with dynamic mass independent on $\Lambda$ (up to
$\log$,s);
\item[(iii)] $\Lambda^{-2}$ (etc.)-orders are irrelevant (though they are
subject to compensation as well);
\item[(iv)] the real dimension in $\Lambda$ is estimated according to the
large-$N_c$ analysis of DSB.
\end{itemize}
The large-$N_c$ approach leads to the following approximation for v.e.v.
of fermion operators,
\begin{equation}
\langle (\bar\psi\psi)^n \rangle \,=\, \biggl(\langle\bar\psi\psi\rangle
\biggr)^n \biggl( 1\,+\,O(1/N_c)\biggr),
\end{equation}
where any number of derivatives is accepted between fermion operators.

V.e.v. of bilinear operator is estimated in the assumption that quarks obtain
a dynamic mass. Namely,
\begin{equation}
\langle\bar\psi \biggl(\frac{\partial^{2}}{\Lambda^{2}}\biggr)^{n} \psi
\rangle\,  \sim  \,\frac{1}{\Lambda^{2n}}
\int \limits_{|p|<\Lambda} \frac{d^{4}p}{(2\pi)^{4}}\,\, tr \frac{p^{2n}}
{\not\! p + i m}\,\sim\, N_{c} m \Lambda^2 .
\end{equation}
One can see that the inclusion of derivatives
does not suppress the contribution
into the mass-gap equation (compare to Eq.8).

We omit the full classification of effective
vertices relevant in the mass-gap Eq.\\(see$^{10}$) and report only the
minimal structure of extended GN model which admits the polycritical regime,
\begin{equation}
{\cal L} = \bar\psi \not\!\! D \psi +  \frac{1}{4N_{c}\Lambda^2}
\sum_{m,n = 0}^{\infty} a_{mn}\, \bar\psi
\biggl(\frac{\partial^2}{\Lambda^2}\biggr)^{n} \psi\cdot \bar\psi
\biggl(\frac{\partial^2}{\Lambda^2}\biggr)^{m} \psi.
\end{equation}
where $a_{mn}$ is a real symmetric matrix. Let us diagonalize this matrix
$\sum a_{mn} f^{(i)}_{n} = \lambda_{i} f^{(i)}_{m},\quad \sum
f^{(i)}_{m} f^{(j)}_{m} = \delta^{(ij)}$ and consider the subspace of
eigenvectors $f^{(i)}$ with non-zero eigenvalues. We def\/ine the vertex
functions $\varphi^{i}(\tau) = \sum^{\infty}_{n = 0} f^{(i)}_{n} \tau^{n}$.

Let us now  introduce the appropriate set
of auxiliary f\/ields $\,\chi_{n}(x) \sim const$ and
develop the mean-f\/ield approach,
\begin{equation}
{\cal L}(\chi) \,=\, \bar\psi \biggl(\not\!\! D +
i M(\chi,\partial^{2})\biggr)\psi
+ N_{c} \Lambda^{2} \sum_{i} \chi_{i}\,\lambda_{i}^{-1}\, \chi_{i}\, .
\end{equation}
Herein the summation is extended over all non-zero eigenvalues
$\lambda_{i}$ (ef\/fective coupling constants). The dynamic mass functional
is a linear combination of vertex functions
\begin{equation}
M(\chi,\partial^{2}) \equiv \sum_{i}
\chi_{i} \varphi^{i}\biggl(\frac{\partial^2}{\Lambda^2}\biggr)\, .
\end{equation}
The corresponding ef\/fective potential in the adiabatic approach can be
derived with the regularization$^{12}$ as in the simple GN model. Then the
generalized mass-gap equation delivers an extremum to ef\/fective
potential and in the $\Lambda^2$-order (fine-tuning) reads,
\begin{eqnarray}
\frac{\partial V_{ef\/f}}{\partial \chi_{i}} = 0 \,\simeq\,
\frac{N_{c}\Lambda^2}{4\pi^2}\, \sum_{j}\,
\biggl(\frac{8\pi^{2}}{\lambda_i}\,\delta_{ij} -
\Phi_{ij} \biggr)\, \chi_{j}\, ,\nonumber\\
\Phi_{ij} = \int\limits_{0}^{1} d\tau \varphi^{(i)}(\tau) \varphi^{(j)} (\tau)
= \sum_{n,m = 0}^{\infty} \frac{f_{m}^{(i)} f_{n}^{(j)}}{m + n + 1}\,.
\end{eqnarray}
It represents the most general condition for subspace of critical
coupling constants. Namely the DSB occurs in the vicinity of
zero-mode subspace for the matrix,$
G_{ij} =  (8\pi^2/\lambda_i )\,\delta_{ij} - \Phi _{ij}$ . By definition the
$(N + 1)$-critical surface (polycritical point)
corresponds to $N$ independent solutions.
We remark that the larger the zero-mode subspace for $G_{ij}$, the larger
may be a symmetry in the DSB phase.

For the clarity we omit the further investigation of mass-gap Eq.
in the general case and proceed to studying the two-channel model.
\section*{\normalsize\bf 4. Two-channel Model around Tricritical Point
and Formation of Two \hspace*{4ex}Resonances}
\hspace*{3ex} This models has three coupling constants $m,n = 1,2$.
The selection of a model with DSB in all scalar channels together with
requirement of scale independence (tricritical condition) leads to the
rigid set of coupling constants.
After appropriate normalization the two-channel model around
tricritical point can be represented by the interaction vertices,
\begin{eqnarray}
{\cal L}_{tricr}\,&=&\, \frac{1}{4N_c\Lambda^2} \biggl[ \lambda^c_1
(\bar\psi\psi)^2 \,+\, 3\lambda^c_2 \biggl(\bar\psi(1 +
2\frac{\partial^2}{\Lambda^2})\psi\biggr)^2 \nonumber\\
&&+\,\frac{8\pi^2}{\Lambda^2}\sum_{i,k} \Delta_{ik}\cdot \bar\psi\phi_i
(- \frac{\partial^2}{\Lambda^2})\psi \cdot
\bar\psi\phi_k (- \frac{\partial^2}{\Lambda^2})\psi\biggr]
\end{eqnarray}
where $\lambda^c_i = 8\pi^2$ and $\phi_1 =1,\; \phi_2 = \sqrt 3 (1 - 2\tau)$.
The dynamic mass function is
\begin{equation}
M(\tau) \,=\, \biggl[\chi_1 + \sqrt 3 \chi_2 (1 - 2\tau)\biggr]
\Theta (1 - \tau).
\end{equation}
The matrix $\Delta_{ik}$ describes the deviation from tricritical point
induced by external forces and is responsible for the mass splitting of
resonances.
Let us display the set of mass-gap Eqs. with accuracy of $O(1/\Lambda^2)$:
\begin{eqnarray}
\chi_1\Delta_{11}\,+\,\chi_2\Delta_{12}\,&=&\, M_0^3 \ln\frac{\Lambda^2}{M_0^2}
\,-\, 6\sqrt 3 \chi_1^2\chi_2 \,-\,18\chi_2^2\chi_1 \,-\, 8\sqrt 3 \chi_2^3 ;
\nonumber\\
\chi_1\Delta_{12}\,+\,\chi_2\Delta_{22}\,&=&\, \sqrt 3
M_0^3 \ln\frac{\Lambda^2}{M_0^2} - 2\sqrt 3 \chi_1^3
 - 18 \chi_1^2\chi_2 - 24\sqrt 3 \chi_2^2\chi_1 - 24 \chi_2^3 .
\end{eqnarray}
where $M_0 = \chi_1 + \sqrt 3 \chi_2 \equiv m_{dyn}$.
This is a set of highly nonlinear Eqs. which cannot be solved analytically.
However we search for solutions around tricritical point which are
smoothly governed by a small deviation $\vert\Delta\vert << \Lambda^2$.
Respectively these solutions are found in the large-$\log$ approximation.

Let us describe the mass spectrum of scalar states near a minimum of $V_{eff}$.
As in the one-channel model the effective kinetic term is defined from the
second variation of effective action,
\begin{equation}
\frac{\delta^2 S_{eff}}{\delta\chi_i(p) \delta\chi_k(p')}
\,\equiv\, \delta(p + p')\biggl(\hat A p^2 \,+\, \hat B\biggr) .
\end{equation}
The scalar state masses are delivered by zeroes of Eq.
\begin{equation}
\Vert\hat A p^2 \,+\, \hat B \Vert \,=\,0 .
\end{equation}
Among solutions one should select out those ones which ensure the positiveness
of the second variation (the minimum) of $S_{eff}$. It can be proven that
such solutions exist and give rise to the physical resonances only  with
masses at $p^2 = - m^2_{\sigma}$ (no tachyons).  Moreover in the vicinity
of tricritical point two scalar resonances are always created when at
least one of coupling constants exceeds its critical value. For an arbitrary
deviation towards overcritical region one discovers two kind of mass spectra:\\
the normal one with lighter mass $m_1^2 \simeq 4 M^2_{dyn}$ (NJL particle)
and with heavier mass $m_2^2 \sim m_1^2 \log \Lambda^2/M^2$;\\
the abnormal one with $ m_1^2 \simeq 6 M^2_{dyn};\quad
m_2^2 \sim m_1^2 \log^{2/3} \Lambda^2/M^2$. Therefore the mass splitting
in general is not completely scale invariant.

Still there is a distinguished direction to drive the mass splitting
independently of $\log\Lambda^2$,
\begin{equation}
\Delta^{scale}_{ik}\,=\, \mu^2 \left( \begin{array}{cc}
1 & \sqrt 3 \\
\sqrt 3 & 3
\end{array} \right),\quad \chi_i \Delta_{ik} \chi_k \,=\,
\mu^2 M^2_0 .
\end{equation}
The lighter scalar state is again a Nambu-GN particle,
$m_1^2 \simeq 4 M^2_{dyn}$ but in the limit $\log\Lambda^2 \rightarrow \infty$
one obtains the constant ratio, $m^2_2 = 20/3 \cdot m_1^2$.

Thus we have started with three arbitrary coupling constants and after
exploiting the cutoff independence of meson mass spectrum we end up with the
single parameter $\mu$ which governs universally the
meson mass splitting. Its possible origin we clarify in the gauged effective
quark model with low-energy gluons.
\section*{\normalsize\bf 5. Two-channel Model with Gluons}
\hspace*{3ex} We develop the gauged two-channel model in a full analogy
with the one-channel case (see Sect.3). Let us follow the large-$N_c$
approach and neglect the $1/\Lambda^2$ orders in the effective action.
As a result modifications concern only the kinetic term of quark fields
and gluon fields turn out not to contribute to higher dimensional vertices
at the orders of $\Lambda^2$ and $\Lambda^0$.  Thereby $V_{eff}$ is extended
with the same one-fermion loop functional of gluon condensates as in the
one-channel model, Eq.10  with replacement
$m_0 \rightarrow M_0 = \chi_1 + \sqrt 3 \chi_2$,
\begin{equation}
\Delta\,V_{eff}(G)\,=\, \Phi (M_0,\,\langle G^n \rangle)\,\simeq\,
\frac{1}{48\pi^2} \ln\frac{M_0^2}{\Lambda^2}
\langle tr(C^a_{\mu\nu})^2 \rangle_0 \,+\, O\biggl(\frac{\langle G^3\rangle}
{M^2_0}\biggr)
\end{equation}
The latter property holds for any number of scalar channels.
Hence the mass splitting driven by gluon condensates happens to be along
the remarkable direction $\Delta^{scale}$ and when $\log \Lambda^2 \rightarrow
\infty$ it takes the same fixed ratio $ m^2_2 =   20/3\cdot m^2_1 $.

The mass gap Eqs. are modified respectively,
\begin{equation}
\frac{\partial V_{eff}}{\partial\chi_i} \,=\,
\frac{\partial V_{eff}}{\partial\chi_i} (G = 0) \,+\,
\frac{\partial \Phi}{\partial M_0} \cdot (1,\,\sqrt 3) \,+\, O(1/\Lambda^2).
\end{equation}
In the large mass expansion the solutions  obey the relations,
\begin{equation}
\chi_1 \,=\, O(1/\Lambda^2);\quad M_0 \,=\,\sqrt 3 \chi_2;\quad
\ln\frac{\Lambda^2}{M^2_0} \,=\, \frac{8}{3} \,+\, \frac{\gamma}{2} .
\end{equation}
Evidently the solution $M_0$ exists for positive gluon condensates and if
$\gamma \simeq 4\div 6$ one obtains $\Lambda /M_0 \simeq 10\div 17$ or when
adopting $M_0 \simeq 300 MeV$ one deals with $\Lambda \simeq (3\div 5) GeV$.
We conclude that the range of applicability of two-channel model is
consistently broader than of the one-channel model.

The scalar mass spectrum is characterized by,
\begin{equation}
m^2_1 \simeq 12\chi^2_2 = 4 M^2_{dyn};\quad m^2_2 \simeq m^2_1\cdot
\frac{20(\gamma - 1)}{3\gamma + 5} .
\end{equation}
Numerically for $M_0 \simeq (300\div 350) MeV$ the scalar state masses are
estmated as $m_1 \simeq (600\div 700)MeV,\quad m_2 \simeq (1.1\div 1.6)MeV$
that is consistent with both the particle phenomenology and with low-energy
expansion in powers of $m^2_{1,2}/\Lambda^2 << 1$.

We notice also that the decay constant (for the lighter meson) proves to be
\begin{equation}
f^2_{\sigma} \,\simeq\, M_0^2 \frac{N_c}{8\pi^2}\,
\biggl(\frac{5}{4}\,+\,\frac{3\gamma}{20}\biggr)_{\gamma = 4\div 6}\,=\,
\frac{N_c}{4\pi^2}\,(1 \pm 0.08) .
\end{equation}
Hence they are positive and less sensitive to the choice of gluon condensate
as compared to the one-channel model.
\section*{\normalsize\bf 6. Conclusion}
\hspace*{3ex} Let us summarize the lessons of studying effective fermion
models with quasilocal vertices.\\
(i)\hspace{3ex} The ef\/fective fermion models with quasilocal interaction
may serve for interpolation of more complicated gauge theories of QCD-type
in the hadronization regime. They contain
the suf\/f\/icient set of phenomenological coupling constants to describe
the inf\/inite spectrum of resonances in the large-$N_c$ approach which is
expected to appear due to conf\/inement.\\
(ii)\hspace{3ex} In the vicinity of tricritical ($n$-critical) point of a
quasilocal fermion model two (respectively $n - 1$) massive scalar states
 occur due to DSB mechanism that may simulate the variety of
 radial excitations of scalar mesons in accordance with the QCD concept.\\
(iii)\hspace{3ex} When effective coupling constants are presribed to provide
the scale invariance of physical parameters (the condition of maximal
polycriticality in open channels) then the position of polycritical point
is uniquely determined in the space of coupling constants.\\
(iv)\hspace{3ex} The minimal sensitivity of mass splitting of scalar states
selects out the particular direction for DSB that leads to considerable
reduction of the number of arbitrary parameters in the two-channel model
$a_{11},\,a_{12},\,a_{22} \rightarrow \mu$. Just in this direction the gluon
medium drives the mass splitting.\\
(v)\hspace{3ex} The inclusion of excited states into the model improves
essentially the predictivity in description of light meson states especially
for models with gluon condensates.\\

We realize that when constructing the realistic quark model in the
hadronization regime one should include vertices with any spin and isospin
structures and derivatives. In this case the selection rule based on the
scale independence may be also applied to related coupling constants.
\section*{\normalsize\bf Acknowledgements}
\hspace*{3ex} We are very grateful to Prof. K. Fujii and Prof. B. L. Reznik
for an opportunity to participate the International School-Seminar '93 -
Hadrons and Nuclei from QCD.
\section*{\normalsize\bf References}
\begin{itemize}
\item[1.] Y. Nambu and G. Jona-Lasinio, {\it Phys. Rev.}
{\bf 122} (1961) 345;\quad
           D. J. Gross and A. Neveu, {\it Phys. Rev.} {\bf D10} (1974) 3235
\item[2.] T. Eguchi, {\it Phys. Rev.} {\bf D14} (1976) 2755;\quad
M. K. Volkov, {\it Ann. Phys. (N.Y.)} {\bf 157}  (1984) 282;\quad
T. Hatsuda and T. Kunihiro, {\it Phys. Lett.} {\bf B145}  (1984) 7;\quad
D. Ebert and H. Reinhard, {\it Nucl. Phys.}, {\bf B271},   (1986) 188;\quad
A. Dhar, R. Shankar and S. R. Wadia, {\it Phys. Rev.} {\bf D3}
 (1985) 3256;\quad
V. Bernard, R. L. Jaf\/fe and U.-G. Meissner, {\it Nucl. Phys.}
{\bf B308} (1988) 753;\quad
   M. Wakamatsu and W. Weise, {\it Z. Phys.- Hadrons and Nuclei}
                 {\bf A331}  (1988)  173.
\item[3.]A. A. Andrianov and V. A. Andrianov, {\it Z. Phys. C - Particles
 and Fields} {\bf 55} (1992) 435;\, {\it Teor. Mat. Phys.}
 {\bf 93} (1992) 67;\quad
             J. Bijnens, C. Bruno and E. de Rafael, {\it Nucl. Phys.}
	     {\bf B390} (1993) 501.
\item[4.] V. A. Miransky, M. Tanabashi and K. Yamawaki,  {\it Mod. Phys. Lett.}
{\bf A4} (1989)  1043; {\it Phys. Lett.} {\bf B221} (1989) 177;\quad
          W. J. Marciano, {\it Phys. Rev. Lett.} {\bf 62} (1989) 2793.
\item[5.] W. A. Bardeen, C. T. Hill and M. Lindner, {\it Phys. Rev.}
{\bf D41} (1990) 1647.
\item[6.] M. Suzuki, {\it Mod. Phys. Lett.} {\bf A5} (1990)  1205;\quad
A. Hasenfratz, P. Hasenfratz, K. Jansen, J. Kuti and Y. Shen,
                 {\it Nucl. Phys.} {\bf B365} (1991)  79.
\item[7.]   K. Wilson and J. Kogut, {\it Phys. Rep.} {\bf 12C} (1974)  75.

\item[8.] M. Veltman, {\it Acta Phys. Polon.} {\bf B8} (1977)  475.
\item[9.] J. Zinn-Justin, {\it Nucl. Phys.} {\bf B367} (1991)  105.
\item[10.] A. A. Andrianov and V. A. Andrianov,
              in {\it Proceedings of Steklov Math. Inst.- LOMI},
{\bf 189/10}, eds. P. P. Kulish and V. N. Popov (Nauka, Leningrad, 1991),
p. 3;\quad{\it Int. J. Mod. Phys.} {\bf 8} (1993) 1981.
\item[11.] B. Rosenstein, B. J. Warr and S. H. Park,
 {\it Phys. Rep.} {\bf 205} (1991)  59 .
\item[12.] A. A. Andrianov and  L. Bonora, {\it Nucl. Phys.} {\bf B233},
 (1984)  232.
\end{itemize}
\end{document}